\documentstyle[eqsecnum,aps,prd]{revtex}
\begin{document}
\thispagestyle{empty}
\begin{flushright}
SUSSEX-AST-95/3-1\\
IEM-FT-104/95\\
gr-qc/9503049\\
March 1995
\end{flushright}
\vskip 1cm
\begin{center}
{\LARGE\bf General Relativity as an Attractor in
\vskip 1mm
Scalar--Tensor Stochastic Inflation}
\vskip 1cm

{\bf Juan Garc\'\i a--Bellido}\footnote{PPARC postdoctoral research
fellow. E-mail: j.bellido@sussex.ac.uk}
\ \ and \ \
{\bf David Wands}\footnote{PPARC postdoctoral research
fellow. E-mail: d.wands@sussex.ac.uk}
\vskip .2mm
{\it Astronomy Centre, School of Mathematical and Physical Sciences,\\
University of Sussex, Brighton BN1 9QH, UK}
\end{center}
\vskip 8mm

{\centerline{\large\bf Abstract}
\begin{quotation}
\vskip -0.4cm
Quantum fluctuations of scalar fields during inflation could determine
the very large-scale structure of the universe. In the case of general
scalar-tensor gravity theories these fluctuations lead to the
diffusion of fundamental constants like the Planck mass and the
effective Brans--Dicke parameter, $\omega$. In the particular case of
Brans--Dicke gravity, where $\omega$ is constant, this leads to runaway
solutions with infinitely large values of the Planck mass. However, in
a theory with variable $\omega$ we find stationary probability
distributions with a finite value of the Planck mass peaked at
exponentially large values of $\omega$ after inflation. We conclude
that general relativity is an attractor during the quantum diffusion
of the fields.
\end{quotation}

\newpage

\section{\label{introduction} Introduction}

One of the most important problems for cosmology is the issue of
initial conditions for the Big Bang \cite{Book}. The first models of
inflation assumed that the universe started in a very hot state that
supercooled in a metastable vacuum, which then decayed to the true
vacuum through a first-order phase transition or just rolled down
through a second-order phase transition \cite{API}. Chaotic inflation
\cite{API} opened up the possibility of starting inflation from a wide
range of initial conditions, including the Planck scale.

Quantum fluctuations in the inflaton field could produce the small
perturbations observed in the background radiation \cite{API,COBE},
from which galaxies later evolved.  Once we include quantum
fluctuations of the scalar fields during inflation, we find that these
can be large and dominate the classical evolution as we approach the
Planck scale \cite{Book}.  As a consequence, the scalar fields
diffuse, very much like a particle in Brownian motion. The universe is
then divided into causally independent inflationary domains, in which
the fields acquire different values.  One of the most fascinating
features of inflation is the process of self-reproduction of the
universe \cite{Book}, by which the values of the fields in some
inflationary domains diffuse towards larger rates of expansion,
producing new domains and so on for ever.  This process might still be
occurring, at scales much larger than our present horizon. The global
behavior of the universe can then be described with the formalism of
stochastic inflation, using probability distributions for the values
of the fields in physical space.

It has recently been shown that there are stationary solutions for
the diffusion of the inflaton in general relativity \cite{LLM}.  A
natural question to ask is whether this picture is still valid as we
approach the Planck era, where quantum fluctuations of the metric
become important \cite{GBL}. Although it is generally assumed that the
dynamics of the universe can be described by general relativity, the
effective theory of gravity might be very different close to the
Planck scale.  So far the only consistent but by no means definite,
since we lack the experimental observations needed to confirm it,
theory of quantum gravity is string theory \cite{GSW}. String theory
contains in its massless gravitational sector a dilaton scalar field
as well as the graviton. The low-energy effective theory from strings
has the form a scalar-tensor theory of gravity \cite{TEGP}, with
non-trivial couplings of the dilaton to matter \cite{CGQ,Polyakov}.
Therefore, it is expected that the description of gravitational
phenomena, and in particular inflation, close to the Planck
scale should also contain this scalar field \cite{Olive,GBQ}.

The string dilaton field can be understood as a Brans--Dicke field
\cite{JBD}, which acts like a dynamical gravitational `constant'.
Jordan--Brans--Dicke theory is the simplest scalar-tensor theory,
with a constant kinetic coupling $\omega$, which is bounded by
primordial nucleosynthesis \cite{PNS} and post-Newtonian experiments
\cite{TEGP} to be $\omega > 500$. String theory predicts $\omega = -1$
in ten dimensions \cite{GSW}. However, the low energy effective value
of $\omega$ depends on the unknown details of the compactification
mechanism and supersymmetry breaking \cite{CGQ}. In general, one would
expect a functional dependence of $\omega$ on the dilaton field
\cite{Polyakov}. Such models were proposed in \cite{HEI,PLB} for solving
the graceful exit problem of extended inflation, and later suggested
to be the generic asymptotic behavior of scalar-tensor theories'
approach to general relativity during the matter dominated era
\cite{Damour,JPMDW}.

In this paper we study the very large scale structure of the universe
assuming that the gravitational interaction is described during
inflation, close to Planck scale, by a general scalar-tensor theory of
gravity, in the context of the stochastic inflation formalism.  In
Brans--Dicke stochastic inflation \cite{GBLL,JGB}, we found runaway
solutions to the diffusion of the dilaton and inflaton fields due to
the fact that the Planck boundary, for generic chaotic potentials, is
a line and the probability distribution does not become stationary, as
occurs in general relativity \cite{LLM}, but slides along this
boundary. As a consequence, the value of Planck mass at the end of
inflation is not well defined (it would depend on new dynamics at
large values of the fields, e.g. quantum loop corrections to the
potential \cite{prep}).  We will study a particular case in which the
Brans--Dicke parameter has a simple pole with respect to the
Brans--Dicke field, which will actually give stationary probability
distributions for the diffusion of the inflaton and dilaton fields in
physical space. We find a probability distribution peaked about
domains that produce general relativity as the effective theory of
gravity at late times and therefore we conclude that it is most
probable to live one of those domains.\footnote{Note that Coleman's
mechanism for the vanishing of the cosmological constant in the
context of Brans--Dicke theory also predicts general relativity as a
low energy effective theory of gravity \cite{Garay}.} This prediction
seems to be in good agreement with observations.

\section{\label{runaway} Runaway Solutions in Brans--Dicke
Cosmology}

In this section we review the problem of runaway
solutions in Brans--Dicke models of stochastic inflation.
For a detailed analysis see Refs. \cite{GBLL,JGB}. Let us
consider the evolution of the inflaton field $\sigma$ with
a generic chaotic potential in a JBD theory of
gravity with dilaton field $\phi$,
\begin{equation}\label{SP}
{\cal S}=\int d^4x \sqrt{-g} \left[{1\over8\omega} \phi^2 R -
{1\over2}(\partial\phi)^2 - {1\over2}(\partial\sigma)^2 -
V(\sigma)\right]\ ,
\end{equation}
where Planck mass is written in terms of the dilaton field as
$M_{\rm P}^2(\phi) = {2\pi\over\omega} \phi^2$. For generic inflaton
potentials of the type $V(\sigma) = \lambda\sigma^{2n}/2n$, the
equations of motion for the homogeneous fields in the slow-roll
approximation are \cite{ExtChaot}
\begin{equation}\label{SRE}
{\dot\phi\over\phi} = {H\over\omega}, \hspace{1cm}
{\dot\sigma\over\sigma} = - {n\over2}\,{H\over\omega}\,{\phi^2\over
\sigma^2}, \hspace{1cm} H = \left({2\omega\lambda\over3n}\right)^{1/2}
\,{\sigma^n\over\phi}\ .
\end{equation}
They correspond to circular motion along the lines
$\phi^2 + {2\over n}\sigma^2 = $ constant, as shown in Fig.~1,
with the angular variable $z = \sqrt{n/2}(\phi/\sigma)$
evolving as \cite{JGB}
\begin{equation}\label{DDZ}
\dot z = {Hz\over\omega}\left(1+z^2\right)\ .
\end{equation}
The end of inflation occurs for $|\dot H| = H^2$,
corresponding to the line $z_e^2 \simeq \omega/2$, where the
slow-roll approximation for the inflaton field breaks down,
while the slow-roll approximation for the dilaton simply requires
$\omega\gg1$. The Planck boundary is the curve  $V(\sigma)
\simeq M_{\rm P}^4(\phi)$, $\ \phi^2 \simeq
(\omega\sqrt\lambda/2\pi\sqrt{2n})\,\sigma^n$. In the simplest
case $n=2$, it is the line $z_p^2 = \omega\sqrt{\lambda}/4\pi$.

Apart from classical motion there are quantum fluctuations
that act on the background fields as stochastic forces and produce
a random motion of those fields, very much like Brownian motion.
The amplitude of quantum fluctuations of $\delta\phi$ and
$\delta\sigma$, whose wavelengths are stretched
beyond the horizon, can be computed in the slow-roll limit, as in
\cite{JGB,STAR}, by solving the linearized perturbation equations
in a de Sitter background ($R = 12H^2 =$ const).
They turn out to be
\begin{equation}\begin{array}{rl}\label{STEP}
&{\displaystyle \delta\phi = \left\langle 4\pi k^3 |u_k|^2
\right\rangle^{1/2} = {H\over2\pi} \ ,}\\[3mm]
&{\displaystyle \delta\sigma = \left\langle 4\pi k^3 |v_k|^2
\right\rangle^{1/2} = {H\over2\pi} \ .}
\end{array}\end{equation}
Quantum fluctuations of the fields act as stochastic forces on
the classical background fields. We will describe the quantum
diffusion of the coarse-grained fields in terms of the
probability distribution $P_p(\sigma,\phi;t)$ to find, at a
given time $t$ in a domain with a given {\it physical} volume,
the fields with mean values $\phi$ and $\sigma$ \cite{GBLL},
\begin{equation}\begin{array}{rl}\label{3HP}
{\displaystyle
{\partial P_p\over\partial t} }=&{\displaystyle
{\partial\over\partial\sigma}\left[{M_{\rm P}^2(\phi)\over4\pi}\,
{\partial H\over\partial\sigma} P_p + {H^{3/2}\over
8\pi^2}\,{\partial\over\partial\sigma}\left(H^{3/2}P_p \right)
\right]} \\[3mm]
+&{\displaystyle
{\partial\over\partial\phi}\left[{M_{\rm P}^2(\phi)\over2\pi}
{\partial H\over\partial\phi} P_p + {H^{3/2}\over8\pi^2}
{\partial\over\partial\phi}\left(H^{3/2}P_p \right)
\right] + 3HP_p }\ ,
\end{array}\end{equation}
where the first and second terms in each bracket correspond to the
classical drift and quantum diffusion, respectively. The last term
takes into account the different quasi-exponential growth of the
proper physical volume in different parts of the universe.  Those few
domains that jump in the opposite direction to the classical
trajectory contribute with a larger physical space and therefore
dominate the proper physical volume of the universe. Such domains
will split into smaller domains, some with lower values of the scalar
fields, some with higher. As a consequence of the diffusion process,
there will always be domains which are still inflating, and this
corresponds to what is known as the self-reproduction of the
inflationary universe \cite{Book}.

Beyond a certain point, the quantum fluctuations $\delta z$
of the fields dominate their classical motion
in a time interval $\Delta t = H^{-1}$ and the
universe enters the self-reproduction regime. In Brans--Dicke
inflation this corresponds to the line $z=z_s$, such that
\begin{equation}
\delta z = \dot{z}\, H^{-1} \ .
\end{equation}
For $n=2$, this is a straight line with
$z_s^4\simeq\omega^3\lambda/12\pi^2\ll1$.
For $z_s<z<z_p$, new domains create
more inflationary domains and so on for ever. (The requirement that
$z_s<z_p$ is guaranteed by the slow-roll condition, $\omega\gg1$.)
In this picture, most of the volume of the
universe today is occupied by regions that are still inflating close
to Planck boundary, while our own causal domain is thought to have
evolved from one of those inflationary domains, down the inflaton
potential, through reheating and into the radiation and matter
dominated eras.

The probability of finding given values of the scalar fields in a
given domain in physical space can be computed, in a first
approximation, with ordinary diffusion equations. In general
relativity, where the Planck boundary corresponds to a certain value
of the inflaton field, it is possible to find stationary solutions to
the quantum diffusion of the inflaton \cite{LLM}. However, in
Brans--Dicke gravity, diffusion occurs in the two-dimensional space
$(\sigma,\phi)$, where the Planck boundary is a line.
For the usual chaotic inflation potentials, the diffusion of the
probability distribution along the Planck boundary is unbounded and
the fields evolve indefinitely along this line \cite{GBLL,JGB} towards
larger and larger values producing what we called runaway solutions.
The diffusion equation for the probability distribution
$P_p(\phi,\sigma,t)$ along the Planck boundary, where $H_P^2 =
(8\pi/3) M_{\rm P}^2(\phi) = (8\pi/3) V(\sigma)^{1/2}$
and diffusion dominates classical drift, can be written as
\begin{equation}\label{FPE}
{\partial P_p\over\partial t} = {1\over8\pi^2}
{\partial\over\partial\sigma}\left(H_P^{3/2}
{\partial\over\partial\sigma}\left(H_P^{3/2} P_p\right)\right)
+ 3H_P P_p\ .
\end{equation}
There are no stationary solutions for generic
chaotic potentials of the type $V(\sigma) \sim \lambda
\sigma^{2n}$. It was argued in Ref.~\cite{GBLL} that a possible way
to stop this runaway behavior of the fields is to introduce a
cutoff in the potential at $\sigma = \sigma_b$ due,
for instance, to quantum corrections.
It is also possible to find stationary solutions within Brans--Dicke
stochastic inflation by considering a non minimal coupling of the
inflaton to the curvature scalar, see Ref.~\cite{prep}.

In this paper we explore another possibility which arises in
more general scalar-tensor gravity theories where the Brans--Dicke
parameter $\omega$ becomes a function of the dilaton field. Such
theories possess the observationally desirable feature of including
the general relativistic behavior in the limit $\omega\to\infty$.
As we shall see, these models can also yield stationary probability
distributions.

\section{\label{vary}Variable omega parameter}

Let us now consider the classical evolution of the inflaton field with
a generic chaotic potential, in the context of an arbitrary
scalar-tensor theory of gravity,
\begin{equation}\label{S}
{\cal S}= \int d^4x \sqrt{-g} \left[ f(\phi) R -
{1\over2}(\partial\phi)^2 - {1\over2}(\partial\sigma)^2 -
V(\sigma)\right]\ .
\end{equation}
Here the parameter $\omega$, a constant in Brans--Dicke theory, becomes a
function of the effective Planck mass
\begin{equation}\begin{array}{rl}\label{PMW}
&{\displaystyle
M_{\rm P}^2(\phi) \equiv 16\pi\,f(\phi)\ ,}\\[3mm]
&{\displaystyle
\omega(f) \equiv {f(\phi)\over2\,[f'(\phi)]^2}\ ,}
\end{array}\end{equation}
where $16\pi f(\phi)$ acts like the Brans--Dicke scalar \cite{JBD}.
The equations of motion of theory (\ref{S}) can be written as
\begin{equation}\begin{array}{rl}\label{XEQ}
\nabla^2\phi\ = &{\displaystyle\!f'(\phi)\, R\ ,}\\[3mm]
\nabla^2\sigma\ = &{\displaystyle\!-\,V'(\sigma)\ , }
\end{array}\end{equation}
\begin{equation}\begin{array}{rl}
{\displaystyle
2f(\phi)\left(R_{\mu\nu} - {1\over2} g_{\mu\nu} R\right)\ =}
&{\displaystyle\! g_{\mu\nu} V(\sigma) + 2 \left(\nabla_\mu
\nabla_\nu - g_{\mu\nu} \nabla^2\right)f(\phi) }\\[3mm]\nonumber
&{\displaystyle + \left(\partial_\mu\phi\partial_\nu\phi -
{1\over2} g_{\mu\nu} (\partial\phi)^2\right) +
\left(\partial_\mu\sigma\partial_\nu\sigma -
{1\over2} g_{\mu\nu} (\partial\sigma)^2\right) }\ .
\end{array}\end{equation}
We can then write the exact equations for the homogeneous fields
in a spatially flat ($k=0$) Friedmann--Robertson--Walker metric as
\begin{equation}\begin{array}{rl}\label{REL}
(2\omega + 3)\left(\ddot f(\phi) + 3H\dot f(\phi)\right)\
+&\!\omega'(f)\dot f(\phi)^2
 = 2 V(\sigma) - {1\over2}\dot\sigma^2\ ,\\[3mm]
\ddot\sigma + 3H\dot\sigma\ =&\!-\,V'(\sigma)\ ,
\end{array}\end{equation}
\begin{equation}\begin{array}{rl}\label{SEQ2}
{\displaystyle f(\phi)\,\left( 6\dot{H} + 12 H^2 \right) \ =}
&{\displaystyle\! 2V(\sigma) - 3 \ddot f(\phi) - 9 H\dot f(\phi)
- {1\over2} \dot\sigma^2 - {1\over2} \dot\phi^2}\ ,\\[4mm]
{\displaystyle f(\phi)\,6H^2}\ =
&{\displaystyle\! V(\sigma) - 6 H \dot f(\phi) +
{1\over2} \dot\sigma^2 + {1\over2} \dot\phi^2 } \ ,\vspace{3mm}
\end{array}\end{equation}
where $H\equiv\dot a/a$, and $a$ is the scale factor. Note that for
$f(\phi) = \phi^2/8\omega$ we recover the usual BD equations.

During inflation, we can write the equations of motion of the
homogeneous fields $\phi$ and $\sigma$, in the slow-roll
approximation, as
\begin{equation}\begin{array}{rl}\label{SEQ}
\dot\phi\ = &{\displaystyle\!4 f'(\phi)\,H
= - {M_{\rm P}^2(\phi)\over2\pi}\,
{\partial H\over\partial\phi}\ ,}\\[4mm]
\dot\sigma\ = &{\displaystyle\!-\,2 f(\phi)
{V'(\sigma)\over V(\sigma)}\,H
= - {M_{\rm P}^2(\phi)\over4\pi}\,
{\partial H\over\partial\sigma}\ ,}\\[4mm]
H^2\ = &{\displaystyle\!{V(\sigma)\over6f(\phi)} }\ .
\end{array}\end{equation}
For the slow-roll solution to be an attractor, $f(\phi)$ must
satisfy the following conditions,
\begin{equation}\begin{array}{rl}\label{SRC}
&f'(\phi)^2\ll f(\phi)\ \Longrightarrow\ \omega(f) \gg 1 \\[3mm]
&{\displaystyle |f''(\phi)|\ll 1\ \Longrightarrow\
{f(\phi)\omega'(f)\over\omega^2(f)} \ll 1}\ .
\end{array}\end{equation}
In addition we have the straightforward generalization to
scalar-tensor gravity of the familiar slow-roll conditions for the
inflaton potential, $\,f(\phi)(V'/V)^2\ll1$ and
$\,f(\phi)|V''/V|\ll1$.

The end of inflation in our theory occurs when $|\dot H| = H^2$, or
$\,V = \dot\sigma^2 + \dot\phi^2 + 3\ddot f(\phi) + 3 H\dot
f(\phi)$. For the generic chaotic potential $V(\sigma) =
{1\over4}\,\lambda\sigma^4$, the end of inflation corresponds, in the
slow-roll approximation, to $\sigma_e \simeq M_{\rm
P}(\phi_e)/\sqrt\pi$.  Note that in our theory, inflation ends when
the inflaton field starts oscillating around the minimum of its
potential, while the dilaton field becomes essentially constant. On
the other hand, the Planck boundary is approximately given by
$V(\sigma) = M_{\rm P}^4(\phi)$ \cite{GBL}, or
$\sigma_p\simeq\sqrt2\,\lambda^{-1/4} M_{\rm P}(\phi_p)$.

The quantum fluctuations in a variable $\omega(f)$ model remain those
of Eqs.~(\ref{STEP}) since they were calculated in the slow-roll
limit, which should still hold here. The variable $\omega(f)$ then
only affects how the diffusion of $\phi$ is reflected in $f(\phi)$,
and thus in the Hubble rate $H$. Along the Planck boundary the rate
of expansion is given by $H_{\rm P}=(8\pi/3)M_{\rm P}^2(\phi)$ and
thus will still diffuse towards larger values of $M_{\rm P}$.

A simple form to consider is $\omega(f) = \omega_0 + \omega_m f^m$
\cite{HEI}, which satisfies the slow-roll conditions for
$\omega_0\gg1$ and $\omega_m > 0$.  For $\,m>0$ we expect that quantum
fluctuations of the fields will drive the effective value of the
Brans--Dicke parameter to infinity, recovering general relativity in
this limit. However, in general, we will not recover a finite value of
$M_{\rm P}$, {\em i.e.}  there will still be runaway solutions for the
quantum diffusion.  A divergent $\omega$ is not sufficient to recover
a stationary probability distribution. However, in the next section we
analyze a scalar-tensor theory that has an upper bound on the Planck
mass which does produce both a stationary distribution and a divergent
$\omega$.

\section{\label{model} The model}

In this section we study the classical evolution of the theory
(\ref{S}), for a particular non minimal coupling $f(\phi)$.  We are
interested in functions that acquire a maximum $(f'(\phi) = 0)$ at
some value of $\phi$, which will lead to a singular value of
$\omega(f)$, see Eq.~(\ref{PMW}). We choose a simple case for which
there are analytical solutions,
\begin{equation}\label{FPH}
f(\phi) = f_0\,\sin^2 a\phi\ .
\end{equation}
In this case, the Brans--Dicke parameter behaves as,
\begin{equation}\label{WPH}
\omega(f) = {\omega_0\over1 - f/f_0} =
{\omega_0\over\cos^2 a\phi}\ ,
\end{equation}
where $a^2 = (8\omega_0 f_0)^{-1} \equiv 2\pi G_0/\omega_0$.
Equation~(\ref{WPH}) has a simple pole at $f = f_0$, where we recover
general relativity with a gravitational constant $G_0$.  This type of
pole behavior for the $\omega$ parameter was used in Ref.~\cite{PLB}
to allow for a graceful exit of extended inflation. Note that it
behaves as ordinary Brans--Dicke theory with $\omega=\omega_0$ for
$a\phi\ll1$.

The slow-roll equations of motion (\ref{SEQ}) for the theory
$V(\sigma) = \lambda\sigma^4/4$ become,
\begin{equation}\begin{array}{rl}\label{EQS}
&{\displaystyle
\dot\phi = \left({\lambda\over3\omega_0}\right)^{1/2}
\sigma^2\,\cos a\phi \ ,}\\[3mm]
&{\displaystyle
\dot\sigma = - \left({\lambda\over3\omega_0}\right)^{1/2}
{\sigma\over a} \sin a\phi\ ,}\\[3mm]
&{\displaystyle
H = \left({\lambda\omega_0\over3}\right)^{1/2}
{a \sigma^2\over\sin a\phi}\ .}
\end{array}\end{equation}
We can redefine $u = (\lambda/3\omega_0)^{1/2}\,t/a\ ,\
x = a \sigma\ ,\ y = \sin a\phi\ ,$
under which the equations of motion (\ref{EQS}) become
\begin{equation}\begin{array}{rl}\label{EQM}
&{\displaystyle y' = x^2 (1-y^2)\ ,}\\[3mm]
&{\displaystyle x' = - x y\ ,}
\end{array}\end{equation}
where a prime here denotes a derivative with respect to $u$.
The slow-roll conditions for the dilaton (\ref{SRC}) require
$\omega_0\gg1-y^2\ $ and $\ \omega_0\gg y^2$, which are both
satisfied by $\omega_0\gg1$.

There are exact solutions to the system of equations (\ref{EQM}),
$x^2 - \ln(1-y^2) =$ constant, see Fig.~1. For small values of $y$,
we recover the circular solutions found in the Brans--Dicke case
\cite{JGB}, while in general we have
\begin{equation}\label{SRS}
\omega(y) = \omega(y_i)\,\exp\left(x_i^2 - x^2\right)\ .
\end{equation}
Note that during inflation $\omega$ increases exponentially.
This behavior will be important for the approach to general
relativity.

In the variable $z = y/x$, the end of inflation and
the Planck boundary correspond to\footnote{Note that these are
the same values as in Brans--Dicke theory \cite{GBLL}.}
\begin{equation}\begin{array}{rl}\label{EIP}
&{\displaystyle z_e^2 \simeq {\omega_0\over2}\ ,}\\[3mm]
&{\displaystyle z_p^2 \simeq {\omega_0\sqrt\lambda\over4\pi}\ .}
\end{array}\end{equation}
The simple analytic results for slow-roll inflation in this model
allows us to describe very simply the evolution of various quantities
such as the number of e-foldings from the end of inflation,
\begin{equation}
N = \omega_0 \ln \left[
 {y_e^2\over y^2} \left( {1-y^2 \over 1-y_e^2} \right) \right] \ ,
\end{equation}
which approaches $N \simeq \omega_0\,x^2$ for $x\gg x_e$.

The new feature in this particular model is the
existence of a boundary in the motion of the dilaton.
As it grows towards $y=1$ it slows down and eventually stops
while the inflaton field $x$ ends inflation.
This effect will be critical for the behavior of the probability
distribution, as we will see shortly. It is simply a consequence
of the local maximum in the function $f(\phi)$, and will occur in
a wide range of functions.

\section{\label{stat} Self-reproduction and stationary distributions}

We will now study the onset of the self-reproduction of the
inflationary universe in our model. The bifurcation
line is defined by the values of the scalar fields for which the
maximum of the probability distribution $P_p$ starts increasing,
or equivalently, where the quantum diffusion of the fields
dominates its classical motion in the time interval $H^{-1}$.
The bifurcation line can be written, in the $z$ variable, as
\begin{equation}
z_s^2 \simeq \left({\lambda\omega_0^3\over12\pi^2}\right)^{1/2}\ .
\end{equation}
There will be self-reproduction of the universe for $z_s > z > z_p$.
The probability distribution $P_p(\phi,\sigma,t)$ follows the
diffusion equation (\ref{3HP}), although, as in the Brans--Dicke
case discussed in section \ref{RUNAWAY}, it will very rapidly diffuse
along the angular direction $z$, towards the Planck boundary, which
is an absorbing boundary \cite{LLM,GBLL}.
It is possible to estimate the relative dispersion in $z$ due to
quantum fluctuations,
\begin{equation}\label{DIS}
{\delta z\over z} = {z_s^2\over\omega_0 z^2}\,\left[(1-y^2) +
z^2\right]^{1/2}\ .
\end{equation}
Note that the prefactor $z_s^2/\omega_0\,z^2 < 2/\sqrt{3\omega_0}\ $
for $z>z_p$ so the dispersion must be small for small values of $z$
and decrease as we approach $y = 1$, which strongly suggests that we
can approximate the diffusion as being essentially one dimensional,
along the Planck boundary. The distribution then evolves along this
boundary following Eq.~(\ref{FPE}), towards large values of $\sigma$
until it reaches the line $y = 1$, which acts like an effective cutoff
and the distribution acquires a peak. Diffusion in $\phi$ across the
boundary $y=1$ is automatically identified with values of $y<1$, see
Eq.~(\ref{FPH}). The upper limit on $\sigma$ is then $\sigma =
(2/G_0\sqrt\lambda)^{1/2} \equiv \sigma_b$. At this point, the
parameter $\omega(f)$ becomes infinite, while $f\omega'(f)/w^2 \to 0$,
corresponding to the general relativistic post-Newtonian limit
\cite{TEGP,DWGB}.

Let us compute the shape of the distribution close to the peak, since
that will give us some idea of possible deviations from general
relativity. The diffusion equation (\ref{3HP}) for the probability
distribution in the physical frame, along the Planck boundary
(\ref{FPE}), for $z_p \ll 1$, can be written in the form of a
Schr\"odinger equation,
\begin{equation}\begin{array}{rl}\label{SCH}
&{\displaystyle
- {\partial^2\Psi\over\partial s^2} + W(s) \Psi(s) =
- E \Psi(s) }\ ,\\[3mm]
&{\displaystyle
P_p(\sigma,t) = \exp\left({E\lambda^{3/4}\,t\over12\sqrt{3\pi}}
\right)\,H_P^{-3/2}\,\Psi(s)\ ,}
\end{array}\end{equation}
where $\Psi(s)$ is the wave function, $s^2 = 1/\sigma$, and
$W(s) = - 72\pi/s^2\sqrt\lambda$ is the singular potential.
It is well known from quantum mechanics that there are no regular
solutions of (\ref{SCH}) at $s=0$ for a singular potential
like $-1/s^2$. However, in our case, the boundary at $\sigma =
\sigma_b$ permits the existence of stationary solutions, with
$\Psi(\sigma_b) = 0$. A numerical solution is presented in
Fig.~2. Unfortunately, we do not have an analytic expression for
it, but one can easily write down WKB approximate solutions as
\begin{equation}\begin{array}{rl}\label{WKB}
&{\displaystyle
\Psi(\sigma) \sim \left(1 - {\sigma\over\sigma_\ast}\right)^{-1/4}
\exp\left\{-\sqrt{72\pi\over\sqrt\lambda}
\left(\sqrt{{\sigma_\ast\over\sigma}-1}
- {\rm arcsec}\sqrt{\sigma_\ast\over\sigma}\right)
\right\}\ ,\hspace{1cm} \sigma < \sigma_\ast \ ,}\\[4mm]
&{\displaystyle
\Psi(\sigma) \sim 2\left({\sigma\over\sigma_\ast} - 1\right)^{-1/4}
\!\!\cos\left\{\sqrt{72\pi\over\sqrt\lambda}
\left[\sqrt{1 - {\sigma_\ast\over\sigma}}\!-
\ln\left(\sqrt{\sigma\over\sigma_\ast} +
\sqrt{{\sigma\over\sigma_\ast}-1}\right)\right]
+{\pi\over4}\right\}\ , \ \sigma > \sigma_\ast \ ,}
\end{array}\end{equation}
where $\sigma_\ast \equiv E\sqrt\lambda/72\pi = \sigma_b
(1 - (9\pi\sqrt\lambda/128)^{1/3})$ is the WKB turning point.
This solution has a very sharp maximum close to $\sigma_\ast$ and
an exponential decay for small $\sigma$.

It is now possible to study the way general relativity is
approached as the probability distribution moves towards the
critical point $\sigma_b$.
The value of the variable Brans--Dicke
parameter {\it along} the Planck boundary can be written as
\begin{equation}
\omega_{\rm P}(\sigma) = {\omega_0\over1-(\sigma/\sigma_b)^2}\ ,
\end{equation}
which shows a pole singularity at $\sigma_b$. Thus near the peak
of the probability distribution function we find
\begin{equation}
\omega_{\rm P} (\sigma_\ast)
 = \left({16\over9\pi}\right)^{1/3}\omega_0\,\lambda^{-1/6} \ ,
\end{equation}
which may be quite large for the values of $\lambda\simeq10^{-12}$
required by density perturbations \cite{Book}.
However, the main increase of
$\omega$ comes from the classical evolution of those domains that
started close to the peak of the distribution and later evolved
towards the end of inflation.\footnote{Note that diffusion is
important between the Planck boundary and the self-reproduction
boundary, however one expects that, as in the case of Brans--Dicke
\cite{GBLL}, the peak of the distribution will closely follow the
classical trajectory.} It is possible to compute this
increase using the slow-roll approximate solution (\ref{SRS}).
By the end of inflation, the effective value of $\omega$
coming from domains at the peak of the distribution takes
the expression
\begin{equation}\label{WEND}
\omega_{\rm end}^\ast \simeq \omega_{\rm P}(\sigma_\ast)
 \,\exp \left[ {2\over\omega_0}
  \left({2\pi\over\sqrt\lambda} - 1\right)\ \right] ,
\end{equation}
which is exponentially large, compared to $\omega_0$. We conclude that
stochastic inflation drives the effective value of $\omega$ at the end of
inflation to be exponentially large, with a sharply peaked probability
distribution. It is important to emphasize here that once the
distribution on the Planck boundary is peaked at $\sigma_\ast$, it is
the classical motion towards the end of inflation which is responsible
for the exponential increase of the effective Brans--Dicke parameter.

\section{Conclusions}

Why do the constants of nature take the values we observe? Couplings
range fourteen orders of magnitude; masses are smaller than $10^{-17}
M_{\rm P}$ and range eleven orders of magnitude; the vacuum energy is
smaller than $10^{-120} M_{\rm P}^4$, and so on.  A possible answer,
very popular among particle physicists, is that there is a unique
logically consistent theory of everything, where all fundamental
constants are determined from its vacuum state. Unfortunately, this
state is probably not unique, {\em e.g.} in superstrings it strongly
depends on the compactification mechanism and supersymmetry breaking
\cite{GSW}. On the other hand, quantum cosmology proposes that the
so-called wave function of the universe provides a probability
distribution for all fundamental constants. It is usually studied in
the canonical or Euclidean approach, which has problems of
interpretation related to the choice of measure. In Ref.~\cite{GBL}
it was suggested that the stochastic inflation formalism could provide
a reasonable framework within which to answer these questions. Here an
exponentially large, causally disconnected inflationary domain
replaces a single nucleated universe of Euclidean quantum
cosmology. This formalism proposes that the global measure should be
given by the probability distribution in physical space
\cite{GBL,Vilenkin}, which takes into account the proper volume of the
universe.

Stochastic inflation describes the quantum diffusion of fields close
to the Planck boundary. It uses branching diffusion equations to
derive the probability of finding a given value of the scalar fields
that drive inflation in a given physical proper volume \cite{LLM}.  It
can be analyzed in the context of general relativity or in other
theories of gravity, like scalar-tensor theories, where the
gravitational coupling (Newton's constant in general relativity),
becomes another dynamical field, the Brans--Dicke field.  In the
stochastic picture this leads to different values of the effective
Planck mass in different exponentially large, causally disconnected,
parts of the universe. This picture may be incorporated in a theory
of evolution of the universe \cite{Evolution}, where quantum
fluctuations of Planck mass could act as a mechanism for mutation,
while a selection mechanism establishes that its value should be as
large as possible in order to increase the rate of expansion, and
therefore the proper volume of the universe. Unfortunately, in the
simplest scalar-tensor theory, Brans--Dicke theory, diffusion close to
the Planck boundary leads to runaway solutions where the global volume
of the universe becomes dominated by regions with infinitely large
Planck mass, in conflict with observations unless new dynamics is
introduced into the model \cite{GBLL,JGB,prep}.

In scalar-tensor theories with power-law behavior of the Brans--Dicke
parameter, there are still runaway solutions, but by considering a
scalar-tensor theory with an upper bound on the BD field
(corresponding to a pole in the BD parameter), we have shown that not
only do we recover a stationary probability distribution for the
fields along the Planck boundary (peaked at the maximum allowed value
of the Planck mass), but that the low-energy effective $\omega$
parameter becomes exponentially large, thus recovering the general
relativistic behavior. It is important to emphasize that this result
is expected to be generic in all models involving a maximum value of
the Planck mass, not just the particular model considered here.

This purely quantum diffusion process towards large values of $\omega$
is then reinforced by the subsequent classical evolution of the
inflationary universe during which the Brans--Dicke parameter
exponentially approaches the general relativistic limit. The ability
of infinite $\omega$ to act as an attractor in classical cosmology is
well-known during inflation \cite{PLB} and matter dominated era
\cite{Damour}. What we have presented in this paper is a
quantum process which enables us to attribute a relative
probability of finding a given value of $\omega$ in the
post-inflationary universe. Along with the classical evolution, this
quantum diffusion mechanism predicts an effective theory of gravity
which at late-times is indistinguishable from general relativity.

\section*{Acknowledgments}

The authors are supported by PPARC. They are grateful to John Barrow,
Ed Copeland, Andrew Liddle and Andrei Linde for useful discussions.



\section*{Figure Captions}

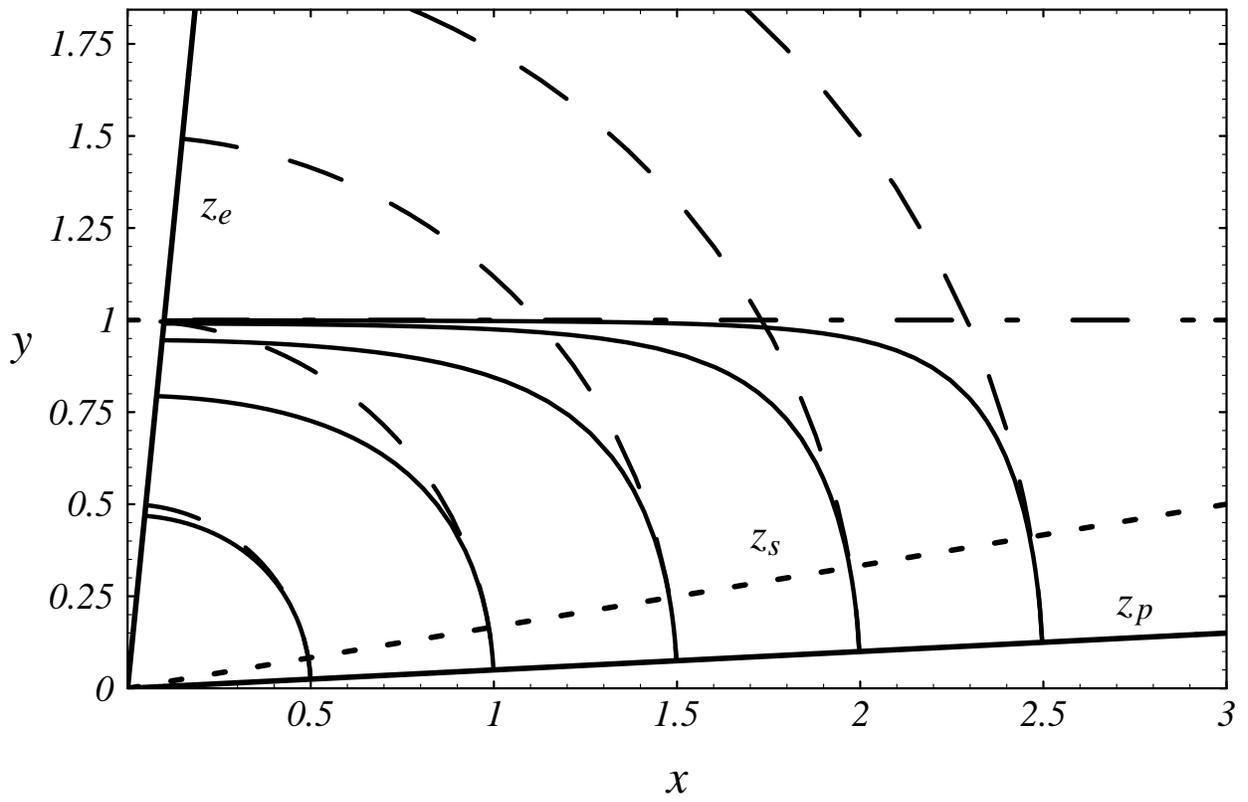
\begin{figure}
\caption{The classical evolution of dilaton and inflaton fields
during inflation in the $(x,y)$ plane, see Eq.~(4.4), for
the scalar-tensor theory defined by (4.1) is represented by the
continuous curves, while the dashed curves correspond to the solutions
in ordinary Brans--Dicke theory. Classical motion starts at Planck
boundary ($z_p$) and ends at the end of inflation boundary ($z_e$),
represented by the thick straight lines, while the dotted line
corresponds to the self-reproduction boundary ($z_s$). The dot-dashed
horizontal line corresponds to the general relativistic limit
$y\to1$. Note that the classical motion in our theory
very quickly approaches that limit.}
\label{fig1}
\end{figure}

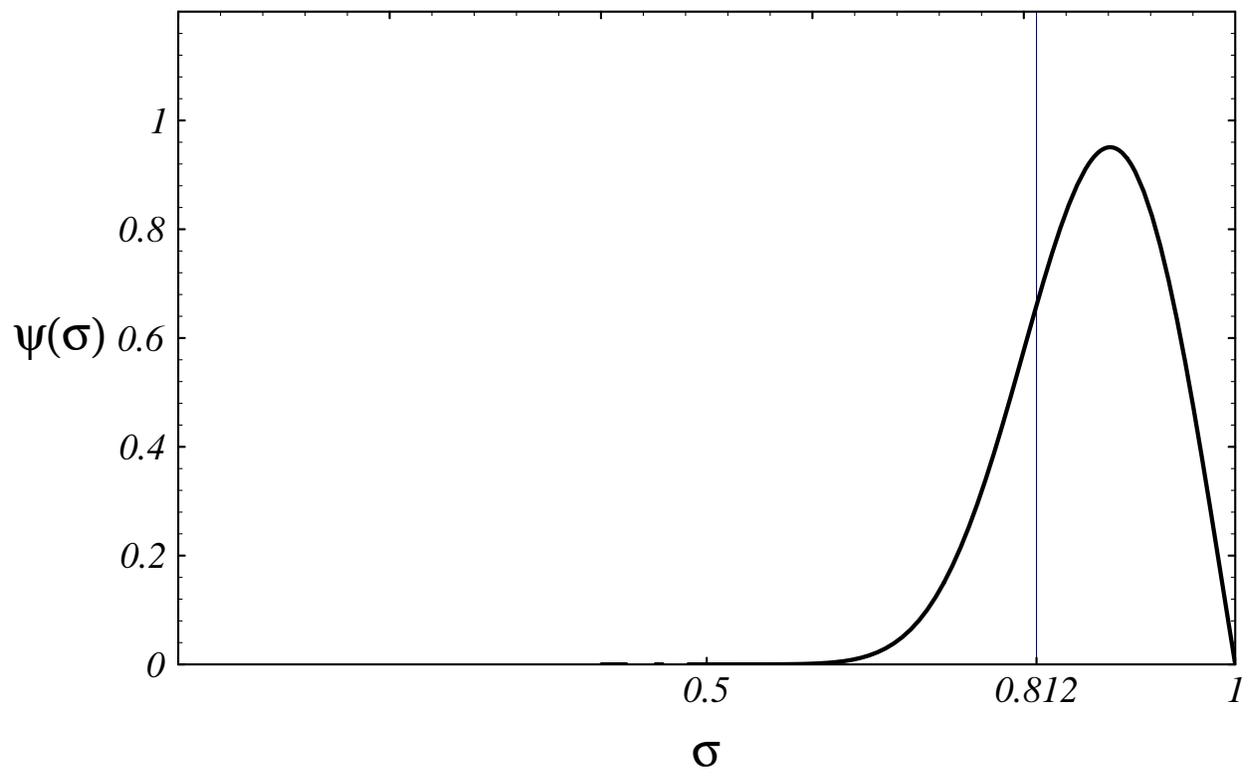
\begin{figure}
\caption{The probability distribution $\Psi(\sigma)$ along the
Planck boundary, for $\sigma_b = 1$ and $\sigma_\ast = 0.812$.
The WKB solutions (5.4) are good approximations to the numerical
result away from the turning point $\sigma_\ast$, marked here
with a vertical line.}
\label{fig2}
\end{figure}

\end{document}